# Correlation between tunneling magnetoresistance and magnetization in dipolar coupled nanoparticle arrays

D. Kechrakos [a] and K. N. Trohidou

Institute of Materials Science, NCSR Demokritos, 15310 Athens, Greece

The tunneling magnetoresistance (TMR) of a hexagonal array of dipolar coupled anisotropic magnetic nanoparticles is studied using a resistor network model and a realistic micromagnetic configuration obtained by Monte Carlo simulations. Analysis of the field-dependent TMR and the corresponding magnetization curve shows that dipolar interactions suppress the maximum TMR effect, increase or decrease the field-sensitivity depending on the direction of applied field and introduce strong dependence of the TMR on the direction of the applied magnetic field. For off-plane magnetic fields, maximum values in the TMR signal are associated with the critical field for irreversible rotation of the magnetization. This behavior is more pronounced in strongly interacting systems (magnetically soft), while for weakly interacting systems (magnetically hard) the maximum of TMR ($H_{\max}$) occurs below the coercive field ($H_c$), in contrast to the situation for non-interacting nanoparticles ($H_c = H_{\max}$) or in-plane fields. The relation of our simulations to recent TMR measurements in self-assembled Co nanoparticle arrays is discussed.

**PACS** : 75.47.-m, 75.75.+a, 75.50.Tt, 75.60.Ej

## I. INTRODUCTION

Intense research activity in the magnetic properties of ordered nanoparticle arrays [1] [2] [3] is motivated on one hand by the potentials of these materials in advancing the magnetic storage density limit to the range of 1Tb/in$^2$ and on the other hand by the basic scientific interest to reveal the underlying mechanism of magnetization reversal in a collection of interacting magnetic nanoparticles. The hexagonal arrangement of self-assembled nanoparticle arrays, rules out the complications introduced by positional randomness in other nanoparticle-based systems (ferrofluids, granular metals) and makes the theoretical analysis simpler. The investigation of the hysteretic behavior and the underlying magnetization reversal mechanism in nanoparticle arrays is a central issue in the research effort on magnetic nanoparticle arrays. The requirements for high packing densities inevitably introduce a new aspect in the magnetization dynamics of these assemblies, namely the collective behavior caused by interparticle interactions. The insulating nature of the surrounding the nanoparticles material, rules out any type of exchange forces between them, because it prevents electron transfer between neighboring nanoparticles. On the other hand, magnetostatic interactions are always present and their effects have been frequently demonstrated so far in experiments on self-assembled arrays. In particular, reduction of the remanence at low temperature [4], increase of the blocking temperature [5][6][7], increase of the barrier distribution width [8], deviations of the zero-field cooled magnetization curves from the Curie behavior [3], difference between the in-plane and normal-to-plane remanence [9] and increase of the blocking temperature with frequency of applied field [10] have been observed and attributed to interparticle magnetostatic interactions. In addition to the experimental works, various numerical studies that focused on the ground state configuration and the hysteresis behavior of dipolar interacting nanoparticle arrays have appeared. The interplay of dipolar interactions and perpendicular anisotropy was shown [11] to induce a reorientation transition below a critical temperature and interaction-induced shape anisotropy of a finite sample controls the magnetization reversal mode. Dipolar interactions were found to decrease the coercive field of magnetic nanoparticle arrays independently of the array topology (square or hexagonal) despite the fact that the ground state configuration is determined by the array topology [12]. The presence of an incomplete second layer with hexagonal structure does not destroy the ferromagnetic (FM) ordering of the ground state [13], while even slight structural disorder within the array destroys that ordering [14]. On the other hand, higher order (quadropolar) magnetostatic interactions were shown to stabilize the long range order of the ground state in a nanoparticle array [15].

---

[a] Electronic mail : dkehrakos@ims.demokritos.gr

Although great theoretical and experimental effort, has been made so far towards the understanding of the magnetic properties of self-assembled nanoparticle arrays, very little work has been done on electronic transport in these systems. In a recent work, Black et al [2] demonstrated that the conductivity of a Co nanoparticle self-assembled film is dominated by spin-dependent tunneling, that leads to a large (~10%) TMR values at low temperature (~$20^0$K). In these experiments a TMR signal with rich structure was observed, that was attributed to the details of the underlying magnetization reversal mechanism. Spin-dependent transport measurements have been previously used as an indirect probe of the micromagnetic structure in spin-valves [16], magnetic tunnel junctions [17], artificial ferromagnetic layers [18], and ferromagnetic rings [19]. The basic idea behind these experiments is that the spin-dependent scattering mechanism leads to a resistivity proportional to the (average) relative orientation of the magnetic moments of separated magnetic regions (either nanoparticles or magnetic domains with different magnetization orientation). Thus resistivity measurements could in principle reveal the underlying magnetic correlations. Indeed, in the above mentioned experiments [16][17][18][19] the underlying micromagnetic structure was efficiently correlated to the magnetoresistance signal. Resistor network (RN) models have been successfully implemented in the interpretation of magnetoresistance measurements in the above mentioned experiments [16][18] and earlier experiments in magnetic granular films [20][21][22]. More recently, Inoue and Maekawa [23] have introduced a resistor network model that interpreted successfully the weak temperature dependent TMR in Co-Al-O granular films. The Inoue - Maekawa model combined the ideas of Herman and Abeles [21] on the electron hopping mechanism in granular metals, according to which the intergranular conductivity decays proportional to the intergranular distance and the charging energy of the grain, with the model of Julliere [24] on spin-dependent transport in magnetic tunnel junctions, according to which the conductivity is proportional to the relative orientation of the magnetizations in the FM layers.

In conductivity measurements in self-assembled Co nanoparticle arrays showed [2] an exponential temperature dependence $\ln G \sim -T^{-1}$ was found, which is characteristic of a thermally-activated tunneling (hopping) process between nanoparticles with negligible size dispersion. Furthermore, contributions to the electric current from a co-tunneling process, were ruled out [2]. Based on these conclusions we suggest that a RN composed of resistors defined according to the Inoue-Maekawa model would be appropriate to study charge transport in self-assembled magnetic nanoparticle arrays.

In this paper we study by numerical simulations the correlations between the micromagnetic structure of hexagonal arrays of dipolar interacting nanoparticles and the tunneling magnetoresistance of the sample. To this end, Monte Carlo simulations of the magnetic configuration at a finite temperature and applied field are performed. The conductivity of the sample is obtained, at equilibrium, by numerical solution of a RN model which incorporates the detailed magnetic configuration.

The remaining of the paper is organized as follows : In Section II, we describe the model of the magnetic structure and the method of simulation. The resistor network model is also described in that section. In Section III we present numerical results and discuss the dependence of the TMR on the applied field, the interparticle distance and the direction of applied field. A discussion of our results and a summary of this work are given in Section IV.

## II. THE MODEL AND THE SIMULATION METHOD

Let us consider $N$ identical spherical particles with diameter $D$ forming a two-dimensional triangular lattice in the *xy*-plane with lattice constant $d \geq D$. The size dispersion of the nanoparticles can be neglected to a good approximation, as for self-assembled samples a very narrow size distribution ($\sigma \approx 5\%$) has been achieved [2]. The particles are single-domain, with uniaxial anisotropy in a random direction and they interact via dipolar forces. The total energy of the system is given as

$$E = g \sum_i \frac{(\hat{S}_i \cdot \hat{S}_j) - 3(\hat{S}_i \cdot \hat{R}_{ij})(\hat{S}_j \cdot \hat{R}_{ij})}{(R_{ij}/d)^3} - k \sum_i (\hat{S}_i \cdot \hat{e}_i)^2 - h \sum_i (\hat{S}_i \cdot \hat{H}) \qquad (1)$$

where $\hat{S}_i$ is the magnetic moment direction (spin) of particle $i$, $\hat{e}_i$ is the easy axis direction, $R_{ij}$ is the center-to-center distance between particles $i$ and $j$. Hats indicate unit vectors. The energy parameters entering Eq. (1) are the dipolar energy $g = \mu^2/d^3$, where $\mu = M_s V$ is the particle moment, the anisotropy energy $k = K_1 V$ and the Zeeman energy $h = \mu H$ due to the applied field $H$. The relative strength of the energy parameters entering Eq. (1), the thermal energy $t = k_B T$ and the treatment history of the sample determine the micromagnetic configuration. However, the transition form single-particle to collective behavior is solely determined by the ratio of the dipolar to the anisotropy energy $g/k = (\pi/2)(M_s^2/K_1)(D/d)^3$. The reported values [3] [5] [9] [10] for fcc or hcp Co nanoparticles are $g/k \sim 0.2 - 0.4 (D/d)^3$, while for the soft ε-Co phase higher values are expected [3] . These values define the range of parameter to be used further on in our simulations.

The magnetic configuration of the nanoparticle ensemble under an applied field $H$ and finite temperature $T$ was obtained by a Monte Carlo simulation, using the standard Metropolis algorithm [25] . At a given temperature and applied field, the system was allowed to relax towards equilibrium using $10^3$ Monte Carlo steps per spin and thermal averages were calculated over the subsequent $10^4$ steps. The results were averaged over 2-10 independent random number sequences corresponding to different realizations of thermal fluctuations. Simulations were performed on a rectangular $L_x \times L_y$ simulation cell with $L_x/d = 16$ *and* $L_y/d = 8\sqrt{3}$. For the simulations of the magnetic structure, we used free boundaries in the *z*-axis and periodic boundaries in the *xy*-plane to avoid undesirable demagnetizing effects due to free poles at the sample boundaries. The dipolar interactions were summed to infinite order in-plane, using the Ewald summation method for a quasi-two-dimensional system [26].

We proceed with the description of the resistor network model employed to study the TMR. For a given micromagnetic configuration $\{\hat{S}_i\}$ of the nanoparticle array we define the conductivity between two nanoparticles *i* and *j* as [2,23]

$$\sigma_{ij} = \sigma_0 \left(1 + P^2 \cos\theta_{ij}\right) \exp\left(-R_{ij}/a - E_c/k_B T\right) \quad (2)$$

where $\sigma_0 = 2e^2/h$ is the conductivity quantum, $P$ is the spin polarization of the conduction electrons, $\cos\theta_{ij} = \left(\hat{S}_i \cdot \hat{S}_j\right)$, $E_c = e^2/2C$ is the activation energy to charge a neutral nanoparticle by addition of a single electron, $C$ is the nanoparticle capacitance relative to its surrounding medium and $a = \sqrt{2m^*U/k_B T}$ is the decay length of the electron wavefunction in the insulating barrier of height *U* relative to the Fermi energy. In all our simulations we assumed $a = d$, as a sufficient requirement to allow charge transfer between neighboring nanoparticles and $P = 0.34$ as an appropriate value for Co nanoparticles [2,27]. Charge conservation on every node of the network implies

$$\sum_j \sigma_{ij} \left(\phi_i - \phi_j\right) = 0 \quad (3)$$

where $\{\phi_i\}$ are the electric potentials. We consider two electrodes attached to the left (cathode) and right (anode) side of the sample along the *x*-axis (Fig. 1) To mimic lithographically grown electrodes we assume that the width of the electrodes (*w*) is smaller than the sample width ($L_y$). Nanoparticles in contact to the electrodes share the same potential with them, thus the boundary conditions are

$$\phi_i = 0 \quad ; \quad i \in C \quad (4)$$

and

$$\phi_i = \phi_0 \quad ; \quad i \in A \quad (5)$$

where $\phi_0$ is the voltage applied across the sample. The boundaries along the *y*-axis are free, namely there is no current flow across the *y*-axis boundary. The effective conductivity of the sample is obtained from the requirement that the total power consumption in the network must be equal to the sum of the power consumptions on all the resistors of the network [28]. Thus, the effective conductivity is given as

$$\sigma = \frac{1}{2\phi_0^2} \sum_{i,j} \sigma_{ij} \left(\phi_i - \phi_j\right)^2 \quad (6)$$

For simplicity we have taken $\phi_0 = 1$, in other words all potentials $\{\phi_i\}$ are scaled by the applied voltage. This assumption does not affect our results since the interparticle conductivities, Eq. (2),

-4-

are voltage independent (Ohmic regime). The set of *N* coupled linear equations in Eq. (3) with the boundary conditions given by Eq. (4) – (5) are solved for the unknown potentials $\{\phi_i\}$ by LU Decomposition [29] and the sample conductivity is obtained from Eq. (6). The result depends obviously of the magnetic configuration, which is used as input to obtain the inter-particle conductivities (Eq. (2)). Consequently, the sample conductivity depends on the applied magnetic field. A thermal average is obtained by averaging the conductivity values over a sequence of equilibrium spin configurations produced by the Monte Carlo algorithm. Finally, the magnetoresistance of the sample is defined by

$$MR(H) = \frac{R(H) - R_s}{R_s} = \frac{\sigma_s - \sigma(H)}{\sigma_s} \quad (7)$$

where $R_s$ and $\sigma_s$ denote the saturation values of the resistivity and conductivity, respectively.

## III. NUMERICAL RESULTS

### A. Dependence of M and TMR on the dipolar strength

*(i) In-plane magnetic field*

We discuss first the variation of field-dependent TMR on the dipolar strength for an in-plane magnetic field. In Fig. 2 we show the lower branch of the hysteresis loop and the corresponding variation of TMR with applied field at a temperature below ($t/k=0.02$) and above ($t/k=0.15$) the blocking temperature ($t_b/k=0.13$). We should mention at this point that the absence of true spin dynamics in the Metropolis MC simulation algorithm causes the lack of a physical time-scale in the algorithm and consequently 'time' is measured in MC steps. The observation 'time' in our simulations is $10^4$ MC steps per spin and corresponds to a physical time of $t_{MC}\sim100$ns [30] for non-interacting nanoparticles. This is much shorter than a typical magnetometry observation time $t_{obs}\sim100$s and consequently a much higher blocking temperature is predicted by our simulations ($T_b \approx K_1 V / 7.7 k_B$) than the typical experimental value ($T_b \approx K_1 V / 25 k_B$). However, Metropolis MC simulations mimic efficiently the role of thermal fluctuations and they reproduce qualitatively the trend of the experimental data as a function of temperature [30].

For non-interacting nanoparticles with random anisotropy the well known result for the remanence at zero temperature ($M_r/M_s = 0.5$) is reproduced. For interacting nanoparticles an increase of the remanence with coupling strength is seen. This trend is dictated by the ferromagnetic character of the dipolar interactions on a hexagonal lattice, which also leads to ferromagnetic long-range ordering at the ground state as has been previously demonstrated by various authors [9,12,13,14,15]. In addition, interactions cause a collective reversal of the magnetic moments under an applied field and as a consequence of that the coercive field decreases with the dipolar strength. The effects of dipolar interactions can also be observed in the MR curves. In particular, the maximum TMR effect ($H_{max}$) occurs at the coercive field ($H_{max}=H_c$) and a clear down shift of the TMR peak position with increasing dipolar strength is observed. The remanent TMR value decreases with interactions, which is explained by the fact that the TMR value is a measure of the misalignment of the magnetic moments in the system. The FM character of dipolar forces in the hexagonal lattice enhances the alignment of the moments at zero field and consequently reduce the corresponding TMR value. Finally, let us comment on the sensitivity of the MR curve, namely the absolute value of the slope with respect to the applied field. In the weak coupling regime ( $g/k \leq 0.2$ ) an *increase* of the sensitivity with increasing dipolar strength is observed, both below and above the coercive field. The same trend is followed by the field-dependent susceptibility in the magnetization curves. In the strong dipolar regime ( $g/k \geq 0.3$ ) however, the sensitivity is reduced again below the non-interacting ($g$ = 0) case but the TMR effect is also drastically suppressed, as is more clearly seen above the blocking temperature (Fig.3), where the coercivity vanishes and the curves become symmetric around the zero field. The underlying physical mechanism that emerges from the above results is that in the weak coupling limit the moments rotate almost incoherently to the applied field with dipolar interactions acting as a perturbation that partially aligns them during rotation. Under reduction of the applied field from negative saturation along the x-axis, dipolar interactions keep the moments aligned along the negative x-axis until the field reaches a large enough positive value. Above this value, reversal of the moments is obtained and the interactions again facilitate the alignment of the moments along the positive x-axis. Thus the TMR sensitivity is enhanced both in the rise and the fall of the TMR curves. With increasing coupling the alignment of the moment during rotation becomes more efficient and eventually, in the strong coupling regime ($g/k\sim$1), dipolar interactions dominate the rotation process and they cause a coherent rotation of the moments. This reversal mode is revealed by an abrupt change of the magnetization orientation at the coercive field and the suppression of the TMR signal.

*(ii) Normal-to-plane magnetic field*

A more dramatic dependence of the magnetic properties on the dipolar strength is expected for an applied field normal to the plane of the array, because dipolar interactions favor the in-plane ordering of the moments, while the applied field drives the moments normal to the plane. The competition between these two orthogonal energy contributions is revealed in the strong dependence of the magnetic properties on the dipolar strength (Fig. 4). With increasing dipolar strength both the remanence and the coercivity are reduced and the hysteretic behavior of the sample is gradually suppressed and eventually lost in the strong coupling regime ($g/k$ ~1.0). Correspondingly, the sensitivity of the TMR curve is constantly reduced with increasing coupling strength and the saturation field increases.

The competition between the in-plane anisotropy, induced by the dipolar interactions and the normal-to-plane applied field is best seen in the infinite coupling limit ($g$ = 1, $k$ = 0) shown in Fig. 5. In this case the system is anhysteretic because the applied field is normal to the easy-plane. At low temperature ($t/g$ = 0.02 ) the magnetization curve increases linearly with the field until the value $h/g \approx 16.5$ when saturation of the moments along the field is achieved. This is a critical field for saturation normal to the plane as it can be verified by the following argument. Consider the low temperature magnetization process. At zero field, the dipoles located on a hexagonal lattice in the xy-plane are in their ground state, namely they are FM ordered along the x-axis (Fig. 1). Upon application of an external field along the z-axis the dipoles rotate coherently in the xz-plane and the moments assume the form $\mu_i = \mu(\sin\theta, 0, \cos\theta)$, where $\theta$ is the azimuth angle of the dipoles. Then the total energy of the system is given by the expression

$$E = -Nh\cos\theta + \frac{1}{2}Ng\left(\sum_i \frac{1}{(r_i/d)^3} - 3\sin^2\theta \sum_i \frac{x_i^2}{(r_i/d)^5}\right) \qquad (8)$$

The critical field ($h_0$) for irreversible rotation of the moments is obtained from the requirement that the first and second derivatives of the total energy are equal to zero. The second sum entering Eq. (8) can be numerically calculated [12] and is equal to $b = +5.51709$, while the term containing the first sum makes a constant contribution to the energy and it is irrelevant to the critical field. After some simple algebra [31] one obtains $h_0/g = 3b \approx 16.551$, which is in very good agreement with the simulation results in Fig. 5.

With increasing temperature ($t/g = 0.5$) the critical field is reduced and the transition to saturation is rounded due to thermal fluctuations. As expected, moment disorder is maximized close to the critical field and consequently the TMR signal shows a peak around this field. This peak is rather weak at very low temperature ($t/g = 0.02$) because the moments rotate coherently. As temperature rises ($t/g = 0.5$) thermal fluctuations of the moments are introduced and a double peak structure of the TMR develops. With further increase of temperature ($t/g = 2.0$) the pair of peaks merges to a single one occurring at zero field. The single-peak behavior of TMR indicates that the system is above the critical temperature for dipolar-induced FM ordering.

## B. Dependence of M and TMR on the magnetic field direction

*(i) Variation of the azimuth angle of the magnetic field*

The variation of the magnetization and TMR at low temperature ($t/k = 0.02$) with the azimuth angle ($\theta$) of the magnetic field for an assembly with weak dipolar coupling ($g/k = 0.1$) is shown in Fig. 6 and for moderate coupling ($g/k = 0.2$) in Fig. 7. The applied field remains in all cases shown in Fig. 6 and Fig. 7 within the *xz*-plane ($\phi=0$). The most important feature in these plots is the *large* decrease of the TMR sensitivity as the magnetic field approaches the z-axis. This trend is clearly seen even in the weak interaction regime (Fig. 6). The strong dependence of the TMR curve on the azimuth angle arises from to the competition between the in-plane anisotropy due to interactions and the off-plane direction of the field. In particular, when an in-plane (x-axis) field is gradually reversed, dipolar interactions decrease the TMR sensitivity by introducing an effective anisotropy barrier to in-plane rotation of the moments, as discussed earlier. Contrary to this behaviour, when the applied field makes an angle with the xy-plane, it acts against the Lorentz field that favors the in-plane alignment of the moments. Consequently, the saturation field is much higher and the TMR sensitivity is reduced. For weakly coupled nanoparticles (Fig.6) the rotation of the moment is governed by the anisotropy energy as deduced from the almost constant value of the coercive field and the TMR peak with the field direction. For moderate coupling (Fig. 7), however, not only the sensitivity decreases more dramatically as the azimuth decreases, but a shift of the coercivity and the TMR peak is seen. An new feature that occurs for moderate coupling (Fig. 7) is that the field corresponding to the TMR maximum ($h_{max}$) can be greater than the coercive field ($h_c$) as occurs for an applied field with azimuth $\theta = 15^0$. The appearance of the TMR peak in hexagonal arrays of nanoparticles at a field higher than the coercive field is in contrast to the commonly met situation in random assemblies of interacting nanoparticles (granular solids), where the maximum signal is observed at the coercive field. As discussed above, in the case of a normal-to-plane magnetic field (Fig. 5), strong dipolar forces can suppress the hysteretic behavior and introduce a critical field at which a TMR peak is observed. Taking this idea one step further, we suggest that that the occurrence of a TMR peak is associated with a critical field rather than the coercive field. We deduce from the TMR data shown in Fig. 7 that for directions close to the normal ($\theta = 0^0$) or close to the plane ($\theta = 90^0$) the critical field is close to the coercive, but the deviation between the two is maximum around $\theta = 15^0$.

*(ii) Variation of the polar angle of the magnetic field*

Dipolar interactions in a hexagonal lattice induce an in-plane anisotropy with three equivalent easy axes that coincide with the symmetry axes of the lattice. The presence of three equivalent easy axes in-plane reduces the anisotropy barriers for in-plane rotation of the moments and render the system weakly anisotropic to in-plane rotations of the magnetization. In Fig. 8 we plot the magnetization and TMR for various values of the polar angle ($\phi$) and for moderate dipolar coupling ($g/k$=0.2). The TMR curves for different in-plane directions of the applied field nearly overlap, underlining the weak anisotropy of the sample to in-plane rotations of the moments. It is only in the strong coupling limit $g/k \sim 1$ (not shown here) that the in-plane anisotropy is dominant and vortices form during reversal of the magnetization, giving rise to steps in the hysteresis curve and jumps in the TMR curve.

## IV. DISCUSSION

Dipolar interaction effects on the MR have been extensively studied experimentally [32] and theoretically [33] in magnetic granular metals, which typically consist of a random assembly of magnetic nanoparticles in a metallic or insulating matrix. Comparing the present results with those for granular metals we could say that the most interesting difference between these two systems, is that *in self-assembled arrays an increase of the field sensitivity to an in-plane fields can be achieved by increasing the surface coverage ( $c \sim (D/d)^2 \sim g^{2/3}$ )*, in contrast to what has been known for random assemblies when the packing density ( $x \sim (D/d)^3 \sim g$ ) is increased. We attribute this feature to the ferromagnetic character of the dipolar interactions on the hexagonal lattice that induce a collective in-plane rotation of the moments. For a normal field, however, the trend of the sensitivity follows that of random assemblies and is reduced with increasing coverage. Given that adjustment of the surface coverage can be experimentally achieved by suitable choice of the capping groups surrounding each nanoparticle [34], we would expect that changes in the TMR signal with variation of surface coverage could be observed.

In recent experiments Black *et al* [2] have measured the TMR effect in self-assembled Co nanoparticle arrays. Small samples of about 10x10 nanoparticles were used to measure the magnetoresistance under an in-plane magnetic field. In these measurements, a rich structure in the field dependent TMR signal was observed and the authors attributed it to the details of the magnetization reversal mechanism. Our simulations with an in-plane magnetic field (Fig. 2) and for g/k~0.1-0.2 correspond to the parameters used in the experiments of Black et *al* [2]. Our results for the hysteresis curves are in good agreement with these experiments. Namely, a remanence value around $M_r/M_s$~0.5 is found and smooth curves are predicted even for interacting samples, in accordance with these experiments. However, no indication for fine structure in the TMR signal is found, at least within our model, that treats in detail the correlations between the magnetic moments. Possibly the observed fine structure in TMR could have a different physical origin than the magnetic correlations between the moments and the associated magnetization reversal mechanism.

In conclusion, we have studied the field-dependence of the magnetization and tunneling magnetoresistance in a hexagonal array of dipolar interacting magnetic nanoparticles with random anisotropy. We showed that for an in-plane applied field, increase of the surface coverage (decrease of interparticle distance) increases the sensitivity of the TMR, through enforcement of the inter-particle dipolar interactions, while with normal-to-plane field the opposite effect is achieved. We demonstrated the occurrence of peaks in the TMR associated with a critical field for the reversible-irreversible transition, that are pronounced for strongly interacting dipolar particles ($g/k > 0.2$) and an applied magnetic field around the normal-to-plane direction. Finally, the TMR signal is more sensitive to variations of the azimuth angle of the field rather than the polar angle. As a final remark, our simulations suggest that magnetoresistance measurements in ordered nanoparticle arrays, as those prepared by self-assembly, could shed light into the magnetization reversal mechanism and facilitate the quantification of the interparticle interactions strength.

## ACKNOWLEDGMENTS
One of the authors (DK) acknowledges discussions with Prof Sir R. J. Elliot and a visiting grant by the Royal Society. This work has been supported by the GROWTH project No. G5RD-CT-2001-00478.

**Figure 1**
Sketch of the nanoparticle array used in our simulations with attached electrodes (C, A) on opposite boundaries along the x-axis. The width of the electrodes shown is $w=10(d\sqrt{3})/2$ and they are coupled only to the outermost nanoparticle of each row.

**Figure 2**
Dependence of the low-temperature ($t/k=0.02$) magnetization and TMR on the interparticle dipolar strength. The magnetic field is applied in-plane along the x-axis. Only the lower hysteresis branch is shown.

**Figure 3**
Dependence of the high-temperature ($t/k=0.15$) magnetization and TMR on the interparticle dipolar strength. The magnetic field is applied in-plane along the x-axis. The blocking temperature for the non-interacting nanoparticles is at ($t_b/k=0.13$ ).

**Figure 4**
Low temperature ($t/k=0.02$) magnetization and TMR for a normal-to-plane (z-axis) magnetic field and various dipolar strengths. Only the lower hysteresis branch is shown.

**Figure 5**
Magnetization and TMR for dipolar coupled isotropic ($k=0$) nanoparticles for temperatures close to zero ($t/g=0.02$), below the ferromagnetic transition ($t/g =0.5$) and above the ferromagnetic transition ($t/g =2.0$). The magnetic field is normal to the plane along the z-axis.

**Figure 6**
Variation of the low-temperature ($t/k=0.02$) magnetization and TMR curves with the direction of the magnetic field relative to the z-axis (azimuth). The field is rotated within the xz-plane. The nanoparticles are weakly coupled ($g/k=0.1$) .

**Figure 7**
Variation of the high-temperature ($t/k=0.02$) magnetization and TMR curves with the direction of the magnetic field relative to the z-axis (azimuth). The field is rotated within the xz-plane. The nanoparticles have moderate dipolar strength ($g/k=0.2$) .

**Figure 8**
Weak variation of the magnetization and TMR curves with the in-plane direction of the applied magnetic field. The temperature is low ($t/k=0.02$) and the nanoparticles are coupled with moderate dipolar strength ($g/k=0.2$).

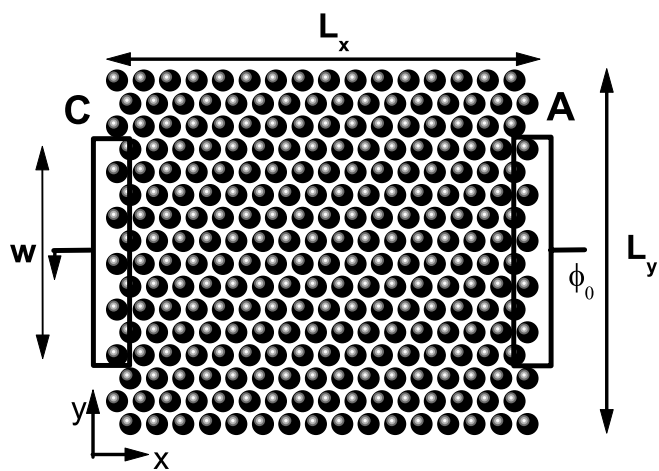

Figure 1.

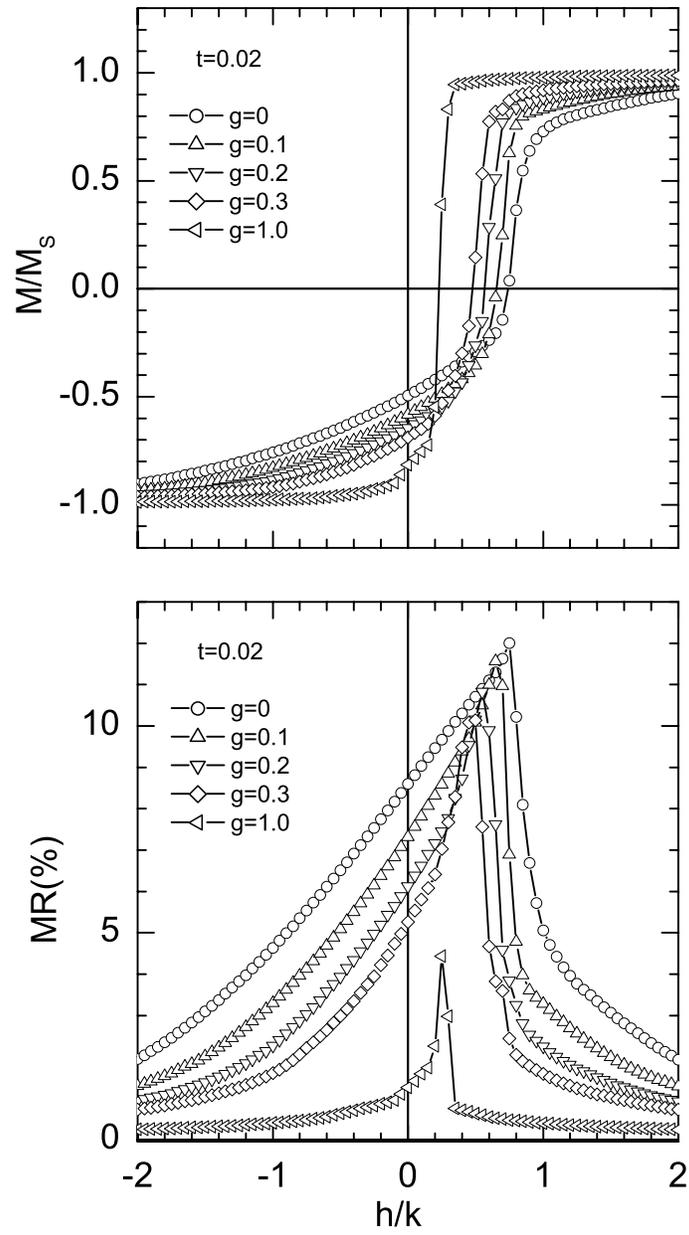

Figure 2

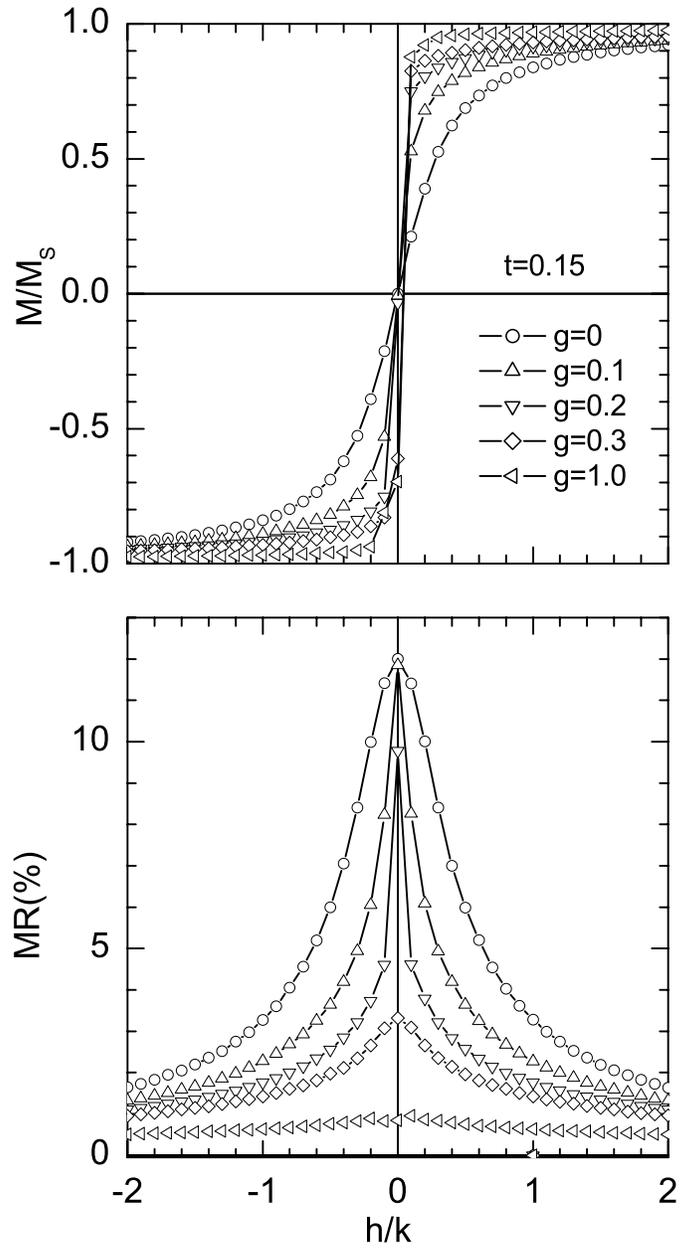

Figure 3.

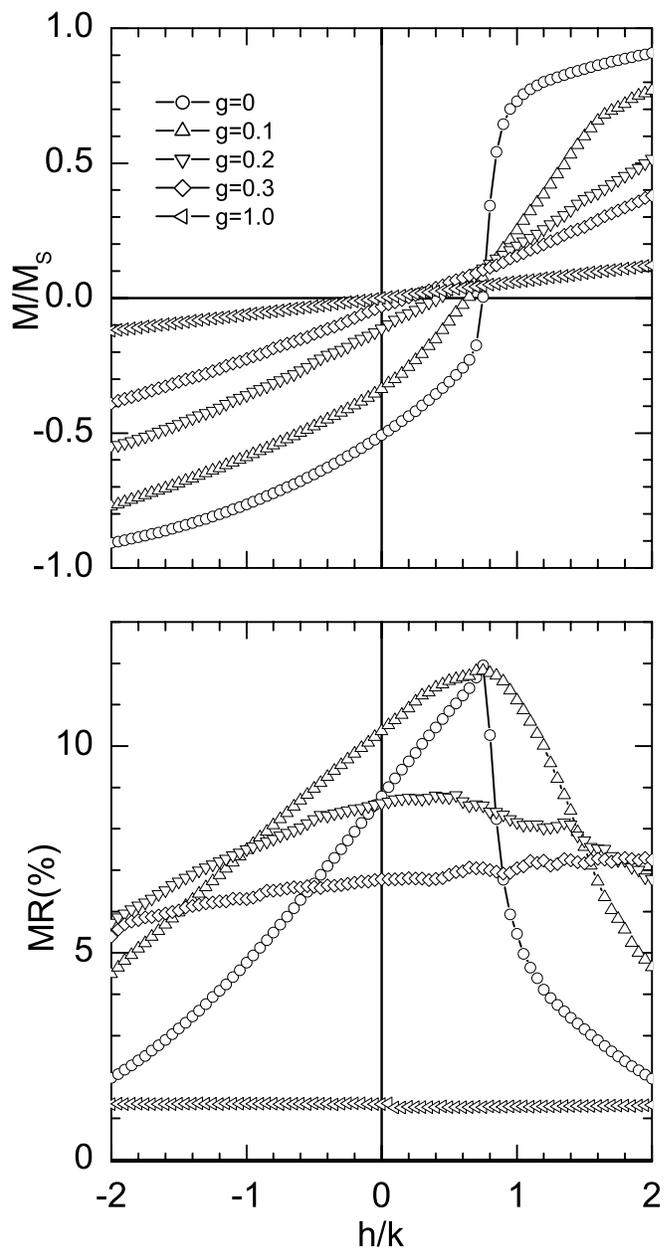

Figure 4

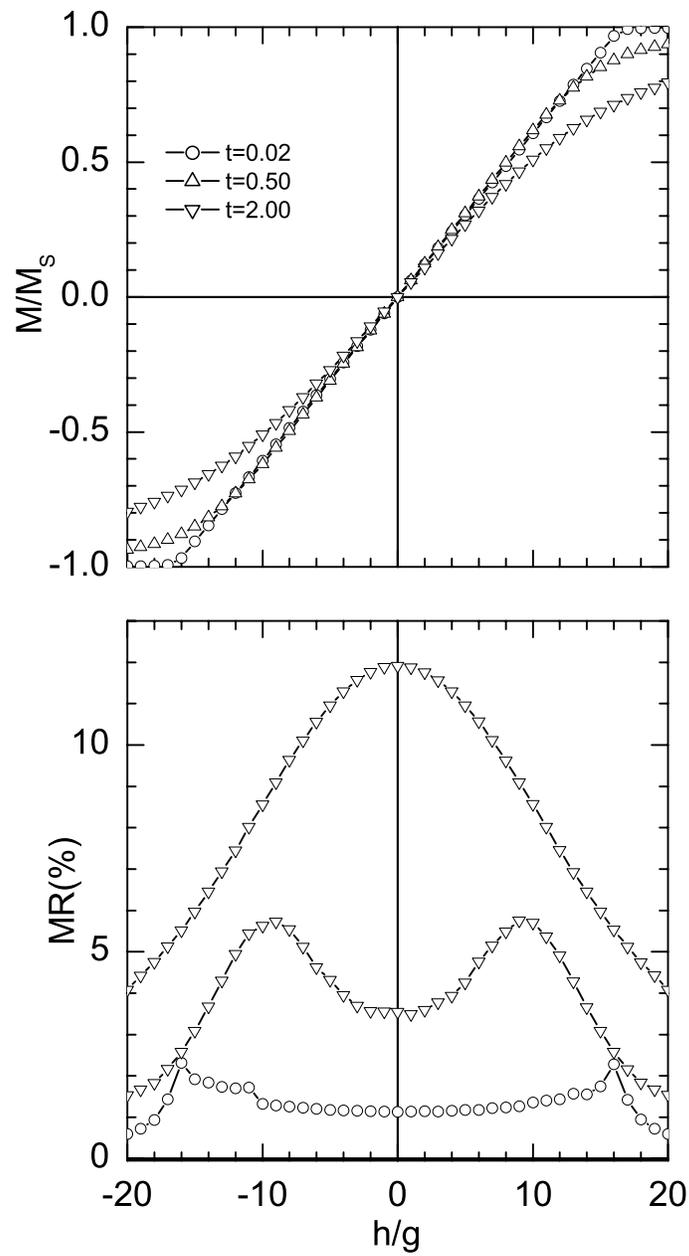

Figure 5

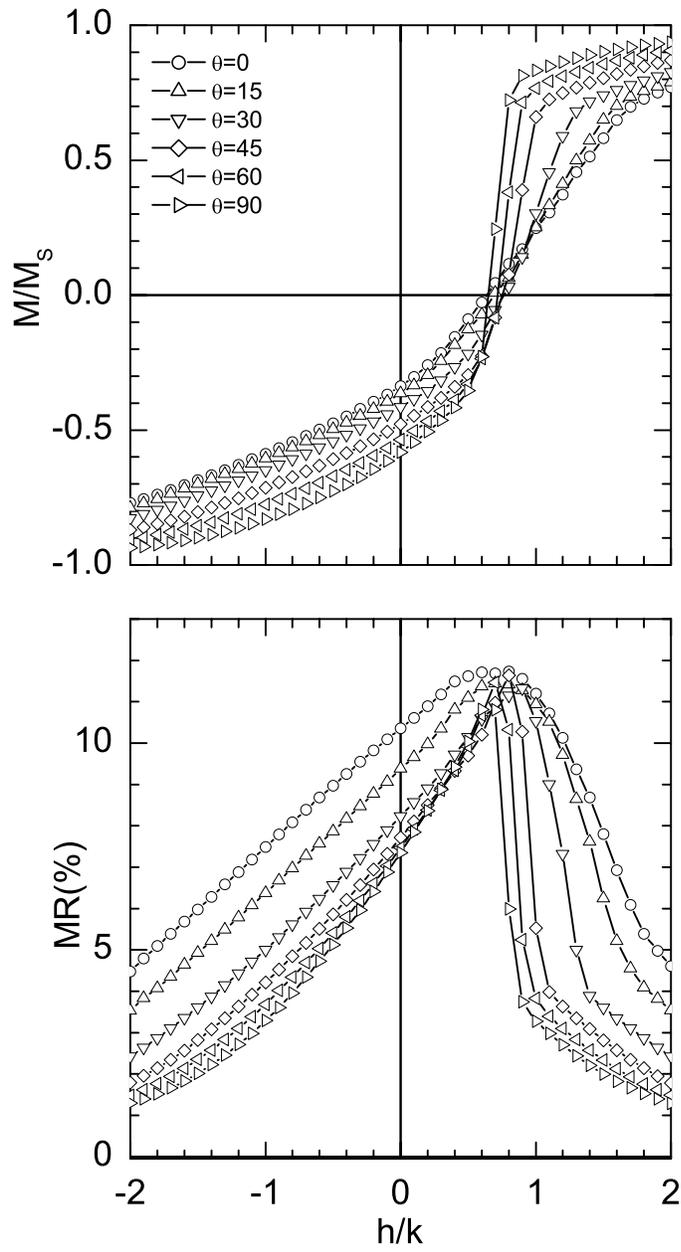

Figure 6

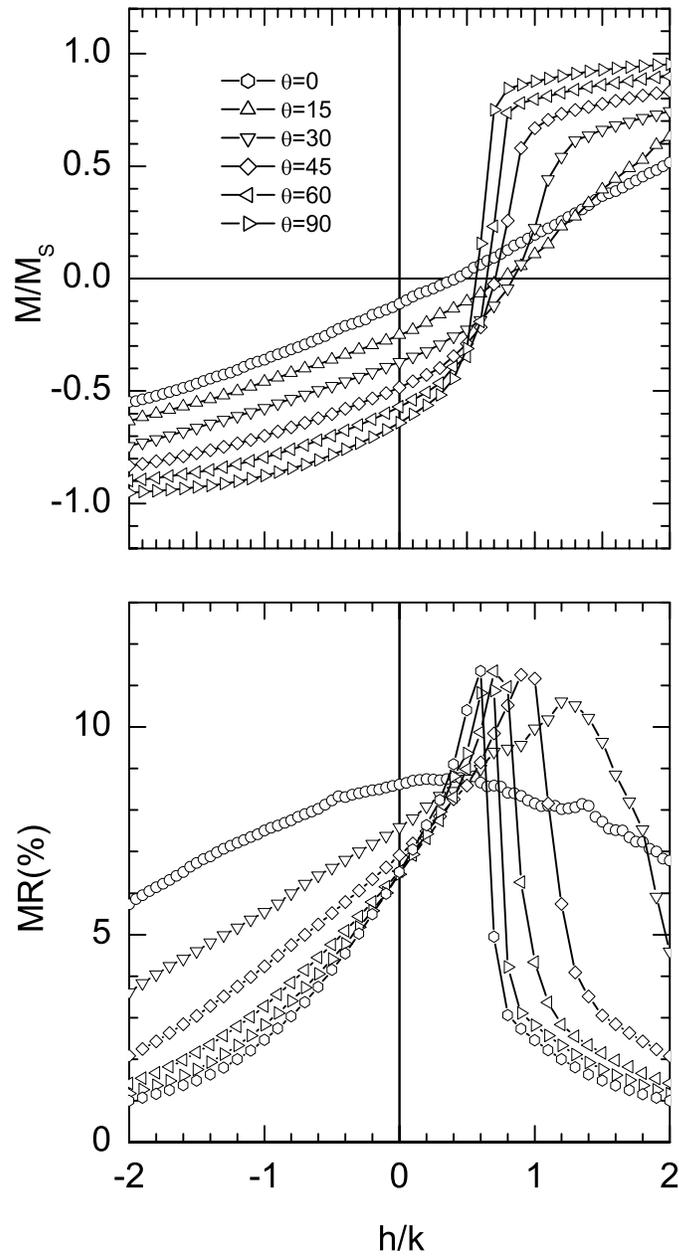

Figure 7

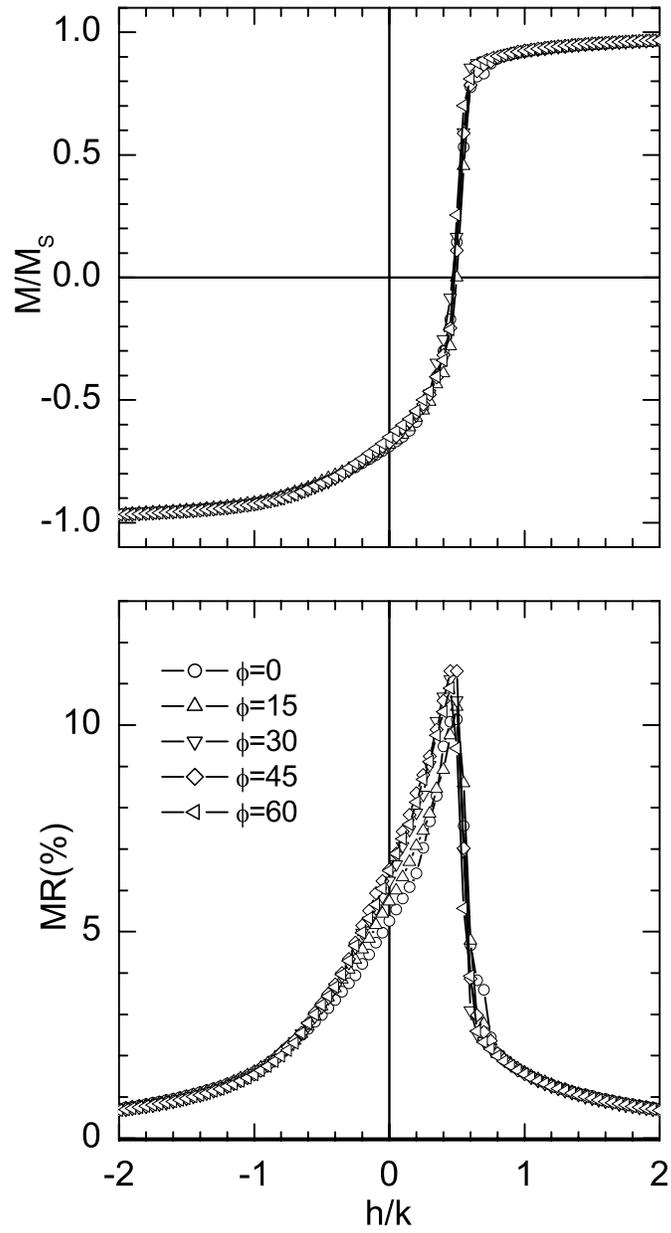

Figure 8